%% file: main.tex
\documentclass[letterpaper, 10 pt, conference]{ieeeconf}
\usepackage{ifthen}
\newboolean{ieeenotice}
\setboolean{ieeenotice}{false}

\usepackage[utf8]{inputenc}
\usepackage[T1]{fontenc}
\usepackage{microtype}
\usepackage{amsmath}
\usepackage{amssymb}
\DeclareMathOperator{\atantwo}{atan2}
\usepackage[english]{babel}
\usepackage[autostyle]{csquotes}
\usepackage[dvipsnames]{xcolor}
\usepackage{graphicx}
\graphicspath{{./pdfs/}{./imgs/}{./fig/}}
\usepackage[backgroundcolor=orange]{todonotes}
\usepackage{siunitx}
\usepackage{tikz}
\usetikzlibrary{arrows,automata,calc,fit,patterns,positioning,backgrounds,shadows,shapes,decorations}
\usepackage{relsize}
\usepackage{ctable}
\usepackage{multirow}
\usepackage{setspace}
\usepackage{paralist}
\usepackage{hyperref}
\usepackage[tight]{units}

\definecolor{paperRed}{rgb}{0.75,0.0,0.0}%
\usepackage{tikz,pgfplots}
\usetikzlibrary{backgrounds,fit,calc,positioning,matrix,decorations.markings,arrows,arrows.meta,fadings,decorations.pathmorphing,shapes.geometric}
\usepgfplotslibrary{colormaps}

\hypersetup{linkcolor=black,citecolor=black,urlcolor=black,colorlinks=true}

\addto\extrasenglish{%
}

\newcommand{\CopyrightNotice}{\hbox
  {\parbox{\textwidth}{\center \textsf{\small $ $\\[-11cm]
    This is a pre-print of an article accepted for publication in \emph{Proceedings of the 30th IEEE Intelligent Vehicles Symposium (IV)}.%\\
  }}}
}

\IEEEoverridecommandlockouts
\title{\Large\textbf{Accuracy Characterization of the Vehicle State Estimation from Aerial Imagery}}
\author{Eduardo Sánchez Morales$^{1}$, Friedrich Kruber$^{1}$, Michael Botsch$^{1}$, Bertold Huber$^{2}$ and Andrés García Higuera$^{3}$
\thanks{$^{1}$Eduardo Sánchez Morales, Friedrich Kruber and Michael Botsch are with the Department of Vehicle Safety, CARISSMA, Technische Hochschule Ingolstadt, Ingolstadt, Germany.}%
\thanks{$^{2}$Bertold Huber is with GeneSys Elektronik GmbH, Offenburg, Germany.}%
\thanks{$^{3}$Andrés García Higuera is with the Department of Electrical, Electronic and Automation Engineering, University of Castilla-La Mancha, Ciudad Real, Spain.}%
}

\begin{document}

\bstctlcite{IEEEexample:BSTcontrol}
\maketitle
\thispagestyle{empty}
\pagestyle{empty}

\ifieeenotice
\CopyrightNotice{}
\fi

\input{architecture.tikz}

\begin{abstract}
Due to their capability of acquiring aerial imagery, camera-equipped Unmanned Aerial Vehicles (UAVs) are very cost-effective tools for acquiring traffic information. However, not enough attention has been given to the validation of the accuracy of these systems. In this paper, an analysis of the most significant sources of error is done. This includes three key components. First, a vehicle state estimation by means of statistical filtering. Second, a quantification of the most significant sources of error. Third, a benchmark of the estimated state compared with state-of-the-art reference sensors. This work presents ways to minimize the errors of the most relevant sources. With these error reductions, camera-equipped UAVs are very attractive tools for traffic data acquisition. The test data and the source code are made publicly available.
\end{abstract}

\section{Introduction and motivation}\label{sec:intro}
%The number of vehicles on the road is steadily increasing every year~\cite{NumberOfCars}~\cite{sousanis2011world}. This rising trend implies that in most parts of the world, there is an interaction between humans and cars. 
Given that cars can harm humans due to malfunction, they have to %fulfill certain safety standards, such as the ISO 26262~\cite{ASIL}. And to comply with such safety standards, vehicles have to 
be tested extensively. Real-world data is a key element for the development, testing and validation of algorithms. %are current trends in the automotive industry. Not only large automotive companies, but research institutes and universities are performing research on these topics as well. One requirement that all these topics have in common, is an accurate vehicle state estimation that can be used as a reference. The only manner to conclusively validate algorithms or sensors, is by using an objective and widely accepted reference. Otherwise, the results obtained are valid only conditionally. 
However, the acquisition of real-world data is resource intensive, which can in turn limit the amount of available data for research.

Given their low cost, versatility and capability of acquiring aerial imagery, camera-equipped UAVs are popular data acquisition tools for automotive purposes. %The term "Unmanned Aerial Vehicle"  applies to all devices that are capable of taking off and remaining airborne for a period of time without a human being physically inside the device. However, for this work, "UAV" is used to referring only to those capable of vertical take-off and hovering, commonly known as "drones". The flight qualities of this type of UAVs, such as payload, static and dynamic stability vary greatly between models; ranging from mini devices with no payload, to systems with payloads of several kilograms. 
Having such video material is highly practical, considering that many vehicle motion models found in the literature describe the movement of cars precisely from this perspective~\cite{schramm2010modellbildung}\cite{abe2015vehicle}\cite{bastow2004car}. Also, since the bird's eye-view is less prone to occlusion problems, it is a highly practical representation for traffic scenarios. Another advantage is the ability of recording various objects with a single device. Some relevant applications of aerial imagery for automotive data acquisition include the traffic flow analysis~\cite{datasky}, the training of machine learning techniques~\cite{8917524} and the validation of autonomous vehicles~\cite{8569682}.

On the other hand, the accuracy validation of camera-equipped UAVs has not received enough attention. When deploying these devices for automotive data acquisition, a series of sensor- and algorithm-related assumptions are made, which translate to inaccuracies. When considering the assumptions individually, the expected error might be negligible. But when considered together, lead to statistically significant errors.

The methodology used in the presented work evaluates the accuracy of camera-equipped UAVs when used for vehicle state estimation. It also clarifies and quantifies the most significant sources of error. The results can then be used as a best practice guide to minimize errors in future works.

This work makes the following contributions:

\begin{inparaenum}[1)]
	It shows
	\item a method for vehicle state estimation when using a camera-equipped UAV as sensor,
	\item an analysis and quantification of the most significant sources of error, and
	\item a benchmark of the estimated vehicle state using state-of-the-art reference sensors.
\end{inparaenum}

This paper is structured in the following manner: first, a brief review on vehicle state estimation by means of cameras is given (\autoref{sec:RelWork}). Then, the methodology for estimating the vehicle state using aerial imagery is detailed (\autoref{sec:Methodology}). Next, the most significant sources of error are analysed and quantified (\autoref{sec:ArrorAnalysis}). Later, the estimated vehicle state is benchmarked with state-of-the-art sensors (\autoref{sec:results}). Finally, the conclusions are presented (\autoref{sec:Conclusions}).

Vectors are represented in boldface and matrices in boldface, capital letters. All the units are given in the International System of Units (SI), unless otherwise specified.

\section{Related Work}\label{sec:RelWork}

The task of using cameras for estimating the vehicle state is not new, and the literature research reveals a wide variety of approaches. An example of a concept using a fixed camera is found on~\cite{cameraVel3}. In that publication, the authors place a camera in the middle of a road and pointed it alongside the traffic flow. After detecting the surface plane, the distance from the vehicles to the camera is calculated. An average velocity is estimated from the calculated distance at each frame.

An example of inclusion of machine learning techniques for velocity estimation can be found in~\cite{cameraVel1}. There, the authors document an algorithm that estimates the velocity and position of vehicles relative to a monocular camera mounted on a moving vehicle. Their method relies on extracting features from videos, and later regress the velocity and position of the vehicles by means of a Multi-Layer Perceptron.

Other works provide already processed video data. Examples of this are the HighD~\cite{highDdataset}, the InD~\cite{inDdataset} and the INTERACTION~\cite{interactiondataset} datasets. The first one provides traffic information for highway. The second and third provide traffic information for urban environments.

The publication of~\cite{cameraVel2} approximates the most to the presented work. There, the authors equipped a test vehicle with a Satellite Navigation (SatNav) system, which allows to measure position and velocity. The main differences from~\cite{cameraVel2} are that in the presented work:
\begin{inparaenum}[1)]
	\item More state variables are benchmarked. Not only the vehicle position and velocity, but the vehicle acceleration, orientation, course over ground and sideslip angle are compared too.
	\item The relief displacement~\cite{HeightPerspective} is corrected. This greatly improves the accuracy of the estimated position. 
	\item The benchmark is done using more accurate sensors. The sensor used in~\cite{cameraVel2} has an accuracy of 20 cm and 0.1 km/h in position and velocity respectively~\cite{vBoxPro}; while the one used in this work has a position and velocity accuracy of 1 cm and 0.03 km/h accordingly. A better sensor accuracy allows a pixel-accurate comparison, which makes the benchmark more relevant.
	\item Test data from a sample video and the source code are publicly available on~\cite{OurDroneCode}. This helps future works to easily generate a measurement vector from aerial imagery for the Kalman Filter (KF), and to fine tune the Kalman gains for specific applications.
\end{inparaenum}
\section{Vehicle State Estimation}\label{sec:Methodology}

The vehicle state is estimated by means of a KF because it allows to calculate state variables by system noise propagation. The detailed process is explained next. 

\subsection{Coordinate Systems}\label{sec:PrevMath}
The coordinate systems used in this work are explained in what follows.

The vehicles move on the Local Tangent Plane (LTP). This plane can be defined in the East-North-Up (ENU) coordinate system, where $x_{\text{LTP}}$ points east, $y_{\text{LTP}}$ north and $z_{\text{LTP}}$ upwards, with arbitrary origin $o_{\text{LTP}}$ on the surface of the Earth. The Local Car Plane (LCP) is defined according to the ISO8855:2011 norm, where $x_{\text{LCP}}$ points to the hood, $y_{\text{LCP}}$ to the driver, $z_{\text{LCP}}$ upwards, with origin $o_{\text{LCP}}$ at the center of sprung mass of the car. For simplification purposes, it is assumed that
\begin{inparaenum}[1)]
	\item the $x_{\text{LCP}}y_{\text{LCP}}$-plane is parallel to the $x_{\text{LTP}}y_{\text{LTP}}$-plane,
	\item the centre of mass of the car is the same as its geometrical centre, and
	\item all sensors mounted on the vehicle measure in the LCP.
\end{inparaenum}
The Pixel Coordinate Frame (PCF) is a bird's-eye image projection of the LTP. It is composed by the mutually perpendicular $x_{\text{PCF}}$ and $y_{\text{PCF}}$ axes, with origin $o_{\text{PCF}}$ in one corner of the image. All quantities expressed in PCF are given in pixels (px).
\subsection{PCF Mapping}\label{sec:GCPs}
The first step for estimating the vehicle state is the mapping of the PCF to the LTP. For this, Ground Control Points (GCPs) are placed on the $x_{\text{LTP}}y_{\text{LTP}}$-plane, in such a way that are visible on the PCF. The middle of the \textit{i-th} GCP is defined in LTP as
\begin{align}
\boldsymbol{g}_{i,\text{LTP}}=
\begin{bmatrix}
x_{i,\text{LTP}}&
y_{i,\text{LTP}}
\end{bmatrix}^\text{T},
\label{eqn:GCPsLTP}
\end{align}
and in PCF as
\begin{align}
\boldsymbol{g}_{i,\text{PCF}}=
\begin{bmatrix}
x_{i,\text{PCF}}&
y_{i,\text{PCF}}
\end{bmatrix}^\text{T}.
\label{eqn:GCPsPCF}
\end{align}
The PCF and the LTP have different scales. The spatial resolution $\alpha$ that relates both scales is calculated from two GCPs by
\begin{align}
\alpha=\frac{\left|{\boldsymbol{g}_{i+1,\text{LTP}}-\boldsymbol{g}_{i,\text{LTP}}}\right|}{\left|{\boldsymbol{g}_{i+1,\text{PCF}}-\boldsymbol{g}_{i,\text{PCF}}}\right|}.
\label{eqn:ScalePCF}
\end{align}
The \textit{i-th} GCP can then be expressed in PCF and in meters by
\begin{equation}
\boldsymbol{g}\prime_{i}=\boldsymbol{g}_{i,\text{PCF}}\cdot\alpha=
\begin{bmatrix}
x\prime_{i}&
y\prime_{i}
\end{bmatrix}^\text{T}.
\label{eqn:GCPinMeters}
\end{equation}
The orientation offset $\xi_{\theta_{\text{image}}}$ from the LTP to the PCF is calculated as 
\begin{align}
\xi_{\theta_{\text{image}}}=\theta\prime_{i}-\theta_{i,\text{LTP}}\text{, with}
\label{eqn:GCPorientedinLTP}
\end{align}
\begin{equation}
\theta\prime_{i}=\atantwo(y\prime_{i+1}-y\prime_{i},x\prime_{i+1}-x\prime_{i})\text{, and}
\label{eqn:GCPorientedinLTP3}
\end{equation}
\begin{align}
\theta_{i,\text{LTP}}=\atantwo(y_{i+1,\text{LTP}}-y_{i,\text{LTP}},x_{i+1,\text{LTP}}-x_{i,\text{LTP}})\text{.}
\label{eqn:GCPorientedinLTP2}
\end{align}
%At least two GCPs have to be used because of possible linear offsets between the PCF and the LTP. The use of two GCPs makes $\theta_{g,i,\text{PCF}}$ and $\theta_{g,i,\text{LTP}}$ relative angles between both GCPs, thus eliminating the effect of linear offsets in the calculation. 
The GCP $\boldsymbol{g}\prime_{i}$ is then rotated as follows
\begin{align}
\boldsymbol{g}\prime\prime_{i}=\boldsymbol{R}\left(\xi_{\theta_{\text{image}}}\right)^\text{T}\boldsymbol{g}\prime_{i},
\label{eqn:GCPorientedinLTP4}
\end{align}
where $\boldsymbol{R}\left(\cdot\right)$ is a 2D rotation matrix.
Finally, the linear offsets $\boldsymbol{\xi}_{\text{d}}$ from the LTP to the PCF are calculated by
\begin{align}
\boldsymbol{\xi}_{\text{d}}=\boldsymbol{g}\prime\prime_{i}-\boldsymbol{g}_{i,\text{LTP}}\text{.}
\label{eqn:GCPorientedinLTP5}
\end{align}
So, a pixel $\boldsymbol{p}_{\text{PCF}}=\begin{bmatrix}
x_{p,\text{PCF}}& 
y_{p,\text{PCF}}
\end{bmatrix}^\text{T}$ on the PCF can be mapped to the LTP by
\begin{equation}
\boldsymbol{p}_{\text{PCF}}^{\text{LTP}}=\left(\boldsymbol{R}\left(\xi_{\theta_{\text{image}}}\right)^\text{T}\left(\boldsymbol{p}_{\text{PCF}}\cdot\alpha\right)\right)-\boldsymbol{\xi}_{\text{d}}\text{.}
\label{eqn:GCPorientedinLTP6}
\end{equation}

\subsection{Measurement Vector Generation}\label{sec:DroneKF:Why}

The next step is to generate a measurement vector from the videos recorded by the UAV and the parameters estimated in \autoref{sec:GCPs}. This requires to process the video sequence for detecting vehicles. The vehicle recognition can be achieved by a variety of machine learning methods that can be found in the literature, such as YOLO~\cite{Redmon.2016}, DroNet~\cite{Kyrkou.2018} or $R^3$~\cite{R3.2018}. The first two output axes aligned bounding boxes; while the third one outputs rotated bounding boxes. The latter greatly improve the state estimation because they allow to calculate the vehicle orientation and a much more precise geometrical center.

The chosen detector for this work is Mask-RCNN~\cite{he2017maskrcnn}\cite{matterport_maskrcnn_2017}. The detailed methodology for processing video material to obtain an adequate bounding box, as well as open source code, can be found on~\cite{KruberFuture}.

The expected input from the UAV is a bounding box per frame of the video sequence that
\begin{inparaenum}[1)]
	\item has 4 sides,
	\item covers all pixels corresponding to the detected shape of the vehicle, and
	\item covers the least amount of pixels possible.
\end{inparaenum}
The corners of the bounding box are given by the intersection of its sides. The \textit{i-th} corner of the bounding box is defined in PCF as 
\begin{align}
\boldsymbol{b}_i=
\begin{bmatrix}
x_{i,\text{PCF}}&
y_{i,\text{PCF}}
\end{bmatrix}^\text{T},
\label{eqn:BoundingBox}
\end{align}
and the bounding box is defined in PCF as 
\begin{align}
\boldsymbol{B}_{\text{PCF}}=
\begin{bmatrix}
\boldsymbol{b}_1&
\boldsymbol{b}_2&
\boldsymbol{b}_3&
\boldsymbol{b}_4
\end{bmatrix}.
\label{eqn:BoundingBox2}
\end{align}
The corners of the bounding box are mapped to the LTP as shown in the Equation (\ref{eqn:GCPorientedinLTP6}) to obtain $\boldsymbol{B}_{\text{PCF}}^{\text{LTP}}$. Let $\boldsymbol{b}_{x}$ and $\boldsymbol{b}_{y}$ be horizontal vectors equal to the first and second rows of $\boldsymbol{B}_{\text{PCF}}^{\text{LTP}}$ respectively. Then, the geometric centre of the vehicle $\boldsymbol{o}_{\text{in}}$ is calculated by
\begin{align}
\boldsymbol{o}_{\text{in}}=
\begin{bmatrix}
\frac{\max\left(\boldsymbol{b}_{x}\right)+\min\left(\boldsymbol{b}_{x}\right)}{2}\\
\frac{\max\left(\boldsymbol{b}_{y}\right)+\min\left(\boldsymbol{b}_{y}\right)}{2}
\end{bmatrix},
\label{eqn:geomcentercalc}
\end{align}
where $\max$ and $\min$ are functions that choose respectively the maximum and minimum value of a vector. 

\begin{figure}%[htb]
	\vspace{2 mm}
	\centering
	\includegraphics[width=0.99\columnwidth]{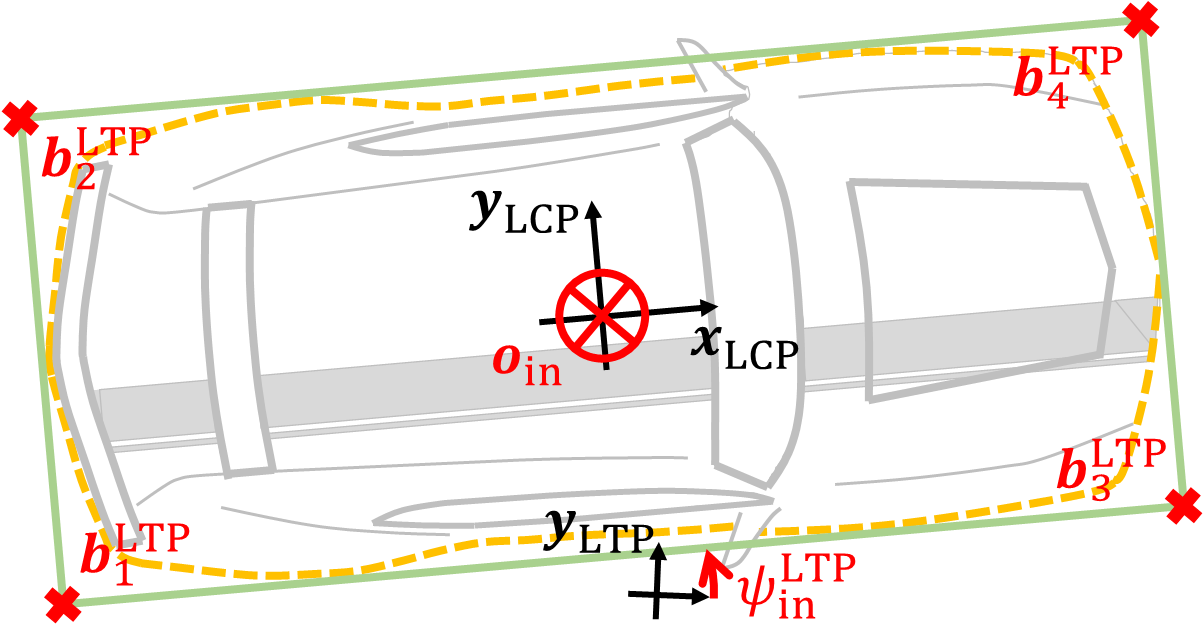}
	\caption{\small{Shown is: the shape of the car as detected by Mask-RCNN with an orange dashed line; the corresponding bounding box with a green solid line; the corners of the bounding box, the geometrical center and orientation of the car with red.}}
	\label{fig:DroneBoundingBox}
	\vspace{-0.5cm}
\end{figure}

The vehicle dimensions are calculated next. Let $\boldsymbol{s}$ be a vector containing the magnitude of the vectors that join the element 1 of $\boldsymbol{B}_{\text{PCF}}$ with all others, so that
\begin{equation}
\boldsymbol{s}\left(1\right)<\boldsymbol{s}\left(2\right)<\boldsymbol{s}\left(3\right)\text{.}
\label{eqn:geomcentercalc2b}
\end{equation}
Then $\left({\boldsymbol{s}\left(1\right)}\cdot{\alpha}\right)$ and $\left({\boldsymbol{s}\left(2\right)}\cdot{\alpha}\right)$ are the estimated width and length of the car in meters. Knowing this, the orientation $\psi^{\text{LTP}}_{\text{in}}$ of the vehicle is given by
\begin{align}
\psi^{\text{LTP}}_{\text{in}}=
\atantwo(y^{\text{LTP}}_{j,\text{PCF}}-y^{\text{LTP}}_{1,\text{PCF}},x^{\text{LTP}}_{j,\text{PCF}}-x^{\text{LTP}}_{1,\text{PCF}})\text{,}
\label{eqn:geomcentercalc3}
\end{align}
where $j$ is the element of $\boldsymbol{B}_{\text{PCF}}$ associated with $\boldsymbol{s}\left(2\right)$. The measurement vector for the KF is then defined as
\begin{align}
\boldsymbol{z_{\text{in}}}=
\begin{bmatrix}
\boldsymbol{o}_{\text{in}}^\text{T}&
\psi^{\text{LTP}}_{\text{in}}
\end{bmatrix}^\text{T}.
\label{eqn:geomcentercalc4}
\end{align}
A graphical representation of the bounding box, vehicle geometric centre and orientation is shown on the \autoref{fig:DroneBoundingBox}. 

\subsection{Kalman Filter}\label{sec:ResultsCompare}

Having the measurement vector, the next step is to estimate the vehicle state. One of the most computationally efficient methods for estimating the \textit{optimum} state of a mobile object, assuming Markovian-Gaussian random processes, is the KF~\cite{Kalman1960}. It also allows to estimate state variables that are not part of the measurement vector by allowing the system noise to propagate. The specifics applicable to this work are described in the following.

The used state vector is defined as
\begin{align}
\boldsymbol{x}=[x_{{\text{car}}},y_{{\text{car}}},v_{x,{\text{car}}},v_{y,{\text{car}}},a_{x,{\text{car}}},a_{y,{\text{car}}},\psi_{{\text{car}}},\dot{\psi}_{\text{car}}]^\text{T},
\label{eqn:KFStateVector}
\end{align}
where $x_{{\text{car}}}$ and $y_{{\text{car}}}$ are the (x,y) coordinates of ${o}_{\text{LCP}}$ in LTP, $v_{x,{\text{car}}}$ and $v_{y,{\text{car}}}$ are the velocities of ${o}_{\text{LCP}}$ along the $x_{\text{LTP}}$ and $y_{\text{LTP}}$ axes, $a_{x,{\text{car}}}$  and $a_{y,{\text{car}}}$ are the accelerations of ${o}_{\text{LCP}}$ along the $x_{\text{LTP}}$ and $y_{\text{LTP}}$ axes, $\psi_{{\text{car}}}$ is the vehicle yaw in LTP, and $\dot{\psi}_{\text{car}}$ is the yaw rate around $z_{\text{LCP}}$.

Since the course over ground $\theta_{{\text{cog}}}$ in LTP of the vehicle is defined as 
\begin{align}
\theta_{{\text{cog}}}=\atantwo(v_{y,{\text{car}}},v_{x,{\text{car}}})\text{,}
\label{eqn:KFEq1}
\end{align}
the sideslip $\beta_{{\text{car}}}$ of the car can then be calculated by
\begin{equation}
\beta_{{\text{car}}}=\theta_{{\text{cog}}}-\psi_{{\text{car}}}\text{.}
\label{eqn:KFEq2}
\end{equation}
%Detailed information about vehicle dynamics and non-tractive driving can be found in~\cite{DrifitingDynamics} and~\cite{schramm2010modellbildung}.
In this work, the sideslip angle is estimated by means of a Linear Kalman Filter (LKF) and Equation (\ref{eqn:KFEq2}). This produces better results than using an Extended Kalman Filter (EKF). This is explained by the fact the sideslip angle is not part of the measurement vector. Using an EKF would imply the estimation of the sideslip angle by noise propagation, whereas Equation (\ref{eqn:KFEq2}) allows a direct calculation.
\section{Analysis of Sources of Error}\label{sec:ArrorAnalysis}
Camera-equipped UAVs are practical, versatile and cost-effective for acquiring traffic information. But their sources of error have not been analysed deeply enough. %The works published so far state the spatial resolution $\alpha$ (as shown in the \autoref{sec:GCPs}) as the expected accuracy. In reality, 
The most significant sources of error are \begin{inparaenum}[1)]
	\item the PCF to LTP Mapping,
	\item the UAV stability,
	\item the car labelling and detection,
	\item the bounding box, and
	\item the sensor synchronization.
\end{inparaenum}
These sources of error are analysed and quantified in the following.
\subsection{PCF to LTP Mapping}\label{sec:GCPError}
\begin{figure}%[htb]
	\vspace{2 mm}
	\centering
	\includegraphics[width=0.99\columnwidth]{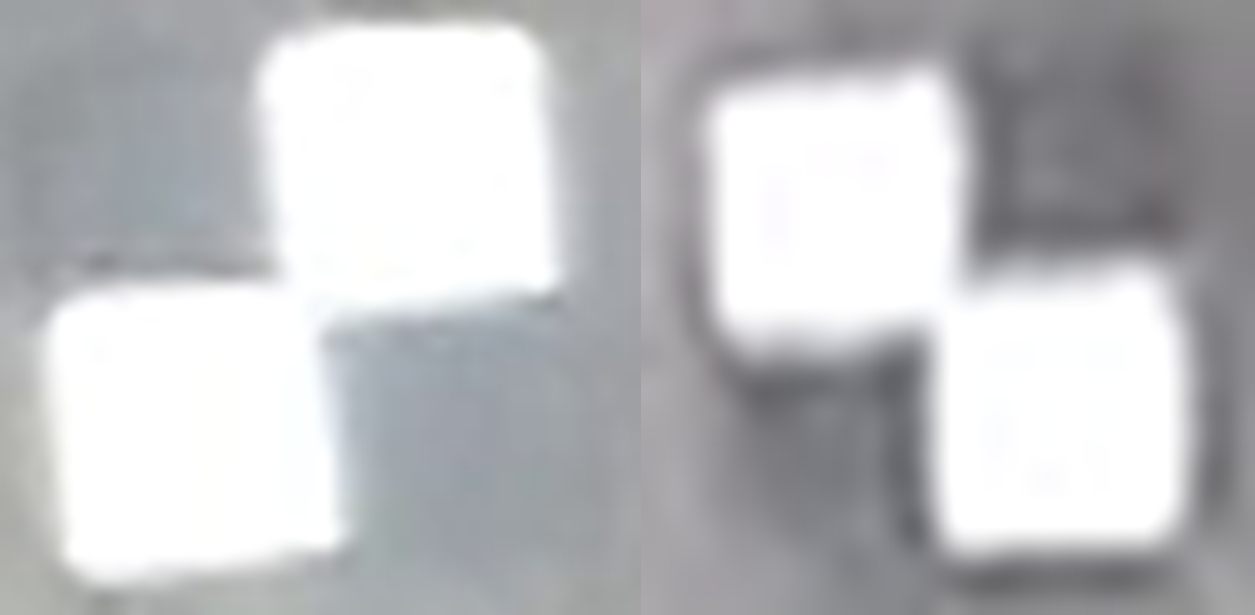}
	\caption{\small{Shown is a GCP as seen on aerial imagery. The UAV hovers at $\unit[50]{m}$ (left) and at $\unit[100]{m}$ (right). The blurring effect of the image prevents to unambiguously assign a pixel to a corner of the GCP.}}
	\label{fig:GCP1}
	\vspace{-0.5cm}
\end{figure}

The literature review shows that the mapping of the GCP to the LTP is usually performed through the use of GCPs. \textit{Ideally}, one point of the GCP can be associated with a specific pixel on the PCF. In reality, this rarely happens. An example of this can be seen on the \autoref{fig:GCP1}. The blurring effect that can be appreciated on the image can be caused by image compression, camera optics or light propagation. It is precisely this blurring effect that prevents to unambiguously associate a point of the GCP with a specific pixel. So, a pixel ambiguity $\zeta=\unit[1]{px}$ in both, the $x_{\text{PCF}}$ and $y_{\text{PCF}}$ axes is not uncommon. So, the $\boldsymbol{g}_{i,\text{PCF}}$ is rewritten as	
\begin{align}
\boldsymbol{g}_{i,\text{PCF}}=
\begin{bmatrix}
x_{i,\text{PCF}}&
y_{i,\text{PCF}}
\end{bmatrix}^\text{T}\pm\zeta.
\label{eqn:GCPError1}
\end{align}

This association ambiguity has an effect on all three parameters that map the PCF to the LTP. % ($\alpha$, $\xi_{\theta_{\text{image}}}$ and $\boldsymbol{\xi}_{\text{d}}$). 
The case of the spatial resolution is analysed first. 

Considering a pixel ambiguity of $\zeta=1$ per axis and squared pixels, the distance between the true and associated positions of a GCP on the PCF can be of $\unit[\sqrt{2}]{px}$. This mis-association causes an error on the spatial resolution. The similarity in percentage $\eta_{\alpha}$ between the seen and the true values of the spatial resolution is calculated by
\begin{align}
\eta_{\alpha}=
\frac{\left|{\boldsymbol{g}_{i+1,\text{PCF}}}-\boldsymbol{g}_{i,\text{PCF}}\right|}{\left|{\boldsymbol{g}_{i+1,\text{PCF}}-\boldsymbol{g}_{i,\text{PCF}}}\right|+2\cdot\sqrt{2\zeta^2}}\text{.}
\label{eqn:GCPError3}
\end{align}
The Equation (\ref{eqn:GCPinMeters}) is then rewritten as
\begin{equation}
\boldsymbol{g}\prime_{i}=\boldsymbol{g}_{i,\text{PCF}}\cdot\alpha\cdot\eta_{\alpha}=
\begin{bmatrix}
x\prime_{i}&
y\prime_{i}
\end{bmatrix}^\text{T}\text{.}
\label{eqn:GCPError4}
\end{equation}
From the Equations (\ref{eqn:GCPError3}) and (\ref{eqn:GCPError4}), it can be deducted that the effect of $\eta_{\alpha}$ increases as $\boldsymbol{g}_{i,\text{PCF}}\to\boldsymbol{g}_{i+1,\text{PCF}}$. In a scenario, where the GCPs are only one pixel apart, the similarity between the seen and true spatial resolution can drop to $26\%$. In another scenario for a Full-HD image, where both GCPs are placed on opposite diagonal corners of the picture, the similarity can drop only to $99.87\%$. %However, depending on the spatial resolution, this can still be statistically significant. 
For example, if the true value of $\left|{\boldsymbol{g}_{i+1,\text{LTP}}-\boldsymbol{g}_{i,\text{LTP}}}\right|$ is $\unit[100]{m}$, a similarity of $\unit[99.87]{\%}$ will rescale it as $\unit[99.87]{m}$, meaning a $\unit[13]{cm}$ difference.

Next, the effect on the orientation offset is analysed. This is done in pixels to decouple errors caused by $\eta_{\alpha}$. To consider the pixel ambiguity, the Equation (\ref{eqn:GCPorientedinLTP3}) is rewritten as 
\begin{align}
\xi_{\theta_{\text{image}}}=\atantwo(\Delta_{y}\pm2\zeta,\Delta_{x}\pm2\zeta)\text{,}
\label{eqn:GCPError5}
\end{align}
where $\zeta$ is multiplied by two because $\xi_{\theta_{\text{image}}}$ is calculated using two GCPs. Similar as with $\eta_{\alpha}$, the effect of the pixel ambiguity increases as $\boldsymbol{g}_{i,\text{PCF}}\to\boldsymbol{g}_{i+1,\text{PCF}}$. In a scenario for a Full-HD image, where the GCPs are one pixel apart, the orientation error $\eta_{\xi_{\theta_{\text{image}}}}$ could reach $116^\circ$. Else, if the GCPs are in opposing diagonal corners of the image, then $\eta_{\xi_{\theta_{\text{image}}}}$ could reach only $0.07^\circ$. 

The orientation error affects the rotation step of the PCF to LTP mapping. So, the effect of orientation error increases as the points to map are farther from the rotation axis. For the \textit{i-th} corner of the bounding box, the mapping error $\eta_{{b}_i}$ due to  the orientation error is given by
\begin{align}
\eta_{{b}_i}=\left|{\boldsymbol{R}\left(\eta_{\xi_{\theta_{\text{image}}}}\right)\boldsymbol{b}_i}-\boldsymbol{b}_i\right|\cdot\alpha\text{.}
\label{eqn:GCPError6}
\end{align}
For example, if the UAV records a Full-HD video while hovering at $\unit[50]{m}$, $\boldsymbol{b}_i\!\!=\!\!
\begin{bmatrix}
1920&
1079
\end{bmatrix}^\text{T}$ and $\eta_{\xi_{\theta_{\text{image}}}}\!\!=\!\!116^\circ$, then $\eta_{{b}_i}\!\!\approx\!\!\unit[124.8]{m}$. Else, if $\eta_{\xi_{\theta_{\text{image}}}}\!\!=\!\!0.07^\circ$, then $\eta_{{b}_i}\!\!\approx\!\!\unit[0.08]{m}$.

The error propagation causes a deviation on the linear offsets as well. The linear offset error $\boldsymbol{\eta}_{{\xi}_{\text{d}}}$ due to the orientation and scaling errors is expressed by	
\begin{align}
\boldsymbol{\eta}_{{\xi}_{\text{d}}}=\left(
\left(\boldsymbol{R}\left(\eta_{\xi_{\theta_{\text{image}}}}\right)^\text{T}\left(\boldsymbol{g}_{i,\text{PCF}}\cdot\eta_{\alpha}\right)\right)-\boldsymbol{g}_{i,\text{PCF}}\right)\cdot\alpha\text{.}
\label{eqn:GCPError7}
\end{align}
In a scenario where the UAV records a Full-HD video while hovering at $\unit[50]{m}$, $\boldsymbol{g}_{i,\text{PCF}}\!\!=\!\!
\begin{bmatrix}
1920&
1079
\end{bmatrix}^\text{T}$, a similarity of $26\%$ and orientation error of $116^\circ$, then $\boldsymbol{\eta}_{{\xi}_{\text{d}}}\approx\begin{bmatrix}
\unit[-2394]{m}&
\unit[-759]{m}
\end{bmatrix}^\text{T}$. Otherwise, with a similarity of $99.87\%$ and orientation error of $0.07^\circ$, then $\boldsymbol{\eta}_{{\xi}_{\text{d}}}\approx\begin{bmatrix}
\unit[-3.83]{m}&
\unit[-0.96]{m}
\end{bmatrix}^\text{T}$. %This means that even with favourable circumstances, alone on the PCF to LTP mapping, a typical error of $\left|\boldsymbol{\eta}_{{\xi}_{\text{d}}}\right|\approx3.95$ pixels can appear for the farthest corner from $o_{\text{PCF}}$.

From the previous, it is deducted that the best way to minimize errors caused by the PCF to LTP mapping, is to locate the GCPs as far from each other as possible. Also, since the direction of the pixel ambiguity is not stochastic, the errors can be compensated by using different combinations of various GCPs.
\subsection{UAV Stability}\label{sec:RegError}
UAVs usually drift while hovering over the LTP. This drift introduces additional scaling, rotation and linear offsets. However, this can be corrected by means of image registration. The registration generally consists of three steps to find correspondences between two images: a feature detector, a descriptor and the matching. The goal of the detector is to find identical interest points under varying viewing conditions. The descriptor is a feature vector, which describes the local area around the point of interest. The SURF \cite{registration1} detector and descriptor is used in this work. The distance between the feature vectors is computed to match the points between two images (matching). An interest point is found on two images when the distance fulfills a certain criterion, e.\,g. a nearest neighbor ratio matching strategy. 
The matches are then fed into the MLESAC algorithm \cite{registration2} to eliminate outliers, i.\,e. incorrect pairs. From the remaining matching pairs, the scaling, rotation and linear offsets are calculated and applied to the video sequence to eliminate the effects of the drift of the UAV.
\subsection{Car Labelling and Detection}\label{sec:RecError}
Another source of error that causes inaccuracies is the labelling and later detection of the vehicle. As stated above, Mask-RCNN is used for detecting vehicles on video sequences. This is done through a process in two steps: the labelling of vehicles for the training of the network, and the use of the trained network to detect vehicles on new video material. To make the detection more robust, neither tires nor mirrors are labeled as part of the vehicle; i.\,e., only the chassis is labeled as vehicle.

\begin{figure}%[htb]
	\vspace{2 mm}
	\centering
	\includegraphics[width=0.99\columnwidth]{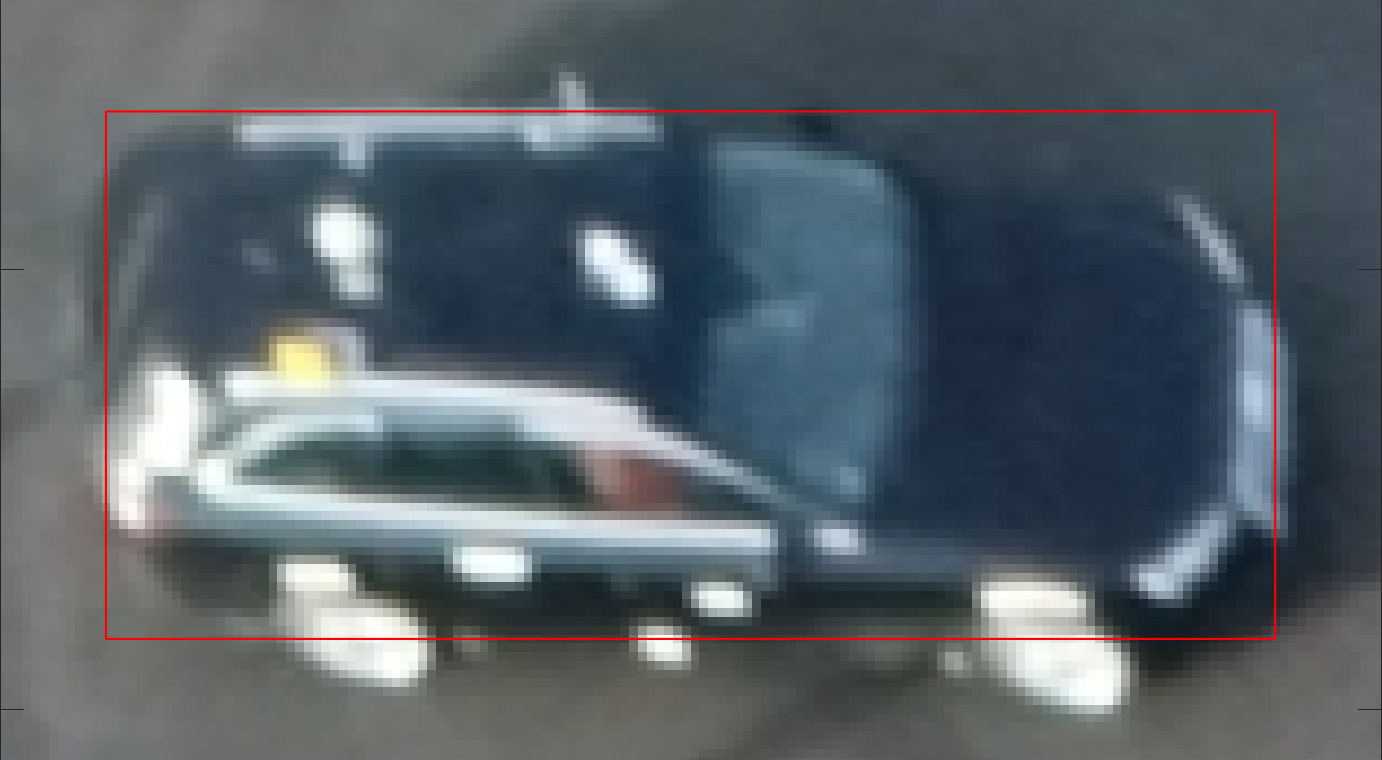}
	\caption{\small{A vehicle with its bounding box.}}
	\label{fig:PerspectiveFar}
	\vspace{-0.5cm}
\end{figure}

Similar as with the GCPs, often the pixels cannot be unambiguously associated with the car. So, a labelling/detection error of \unit[1]{px} or more is common. The effect of this error is propagated to the bounding box. An example of this can be seen on \autoref{fig:PerspectiveFar}.	
\subsection{Bounding Box}\label{sec:BBoxError}
As explained in \autoref{sec:DroneKF:Why}, the measurement vector for the KF is generated from the vehicle bounding box. Since the bounding box is created from the detected shape of the vehicle, this detection is the first source of inaccuracy. It is then realistic to have a pixel ambiguity of $\zeta=\unit[1]{px}$ for each side of the bounding box.

\begin{figure}%[htb]
	\vspace{2 mm}
	\centering
	\includegraphics[width=0.99\columnwidth]{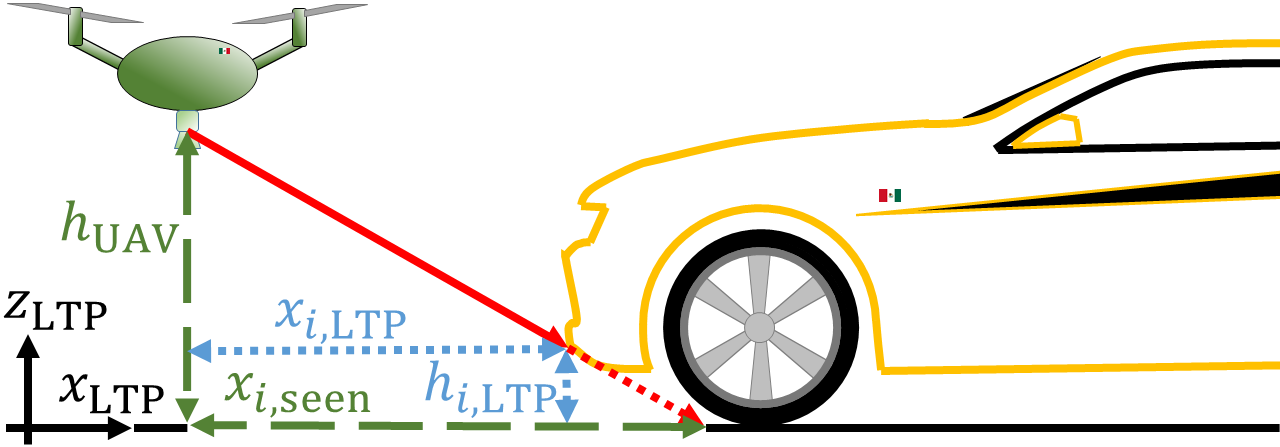}
	\caption{\small{Shown is: the line of sight of the UAV with red solid line, the seen vehicle position with green dashed line and the true vehicle position with blue dotted line.}}
	\label{fig:perspective1}
	\vspace{-0.5cm}
\end{figure}

The next bounding box-related source of error is the perspective. \autoref{fig:DroneBoundingBox} shows an ideal case for the creation of the bounding box, where the complete shape of the vehicle is orthonormal to the UAV, and so can be mapped to the LTP as shown in the \autoref{sec:GCPs}. However, this is not possible because of relief displacement. This optical effect occurs due to camera perspective and the height of the seen objects. It has two consequences for the filmed objects: First, their position shifts away from the principal point (image centre) as their height increases. Second, the dimensions of their bounding box grow if a vehicle side is occluded. A graphical representation of this is shown on \autoref{fig:perspective1}. 

If the height of the object is known, the relief displacement can be corrected. However, because of the combination of camera perspectives, vehicle shapes and poses; the sides of the bounding box could be projections of points with different heights. The corners would then be intersections of projections at different heights. So, the height in LTP of the corners of the bounding box is not known. Even when manually analysing the video material, it is common that pixels cannot be conclusively associated to specific vehicle parts. An example of this is shown in \autoref{fig:PerspectiveFar}.

A statement about the height can be done only for the corner that is closest to the vertical of the drone. Given that only the body is labelled as a vehicle (\autoref{sec:RecError}), it can be deducted that the two sides of the bounding box that are closest to the middle of the picture are tangent to the bottom of the car. So, the intersection of these lines is at this height as well. %It is known to the authors that the bottom might not always be the widest part of the car. Nevertheless, because of stability design, the difference between the widest part of the vehicle and the width of the bottom of the car is negligible. 
Knowing the height of this corner, its relief displacement can be corrected as follows.

Defining the horizontal and vertical resolution of the image as $r_{x}$ and $r_{y}$ respectively, the coordinates in PCF of $\boldsymbol{b}_i$ with respect to the middle of the picture are given by
\begin{equation}
\begin{bmatrix}
x_{i,\text{seen}}\\
y_{i,\text{seen}}
\end{bmatrix}=
\begin{bmatrix}
x_{i,\text{PCF}}-\frac{r_{x}}{2}\\
y_{i,\text{PCF}}-\frac{r_{y}}{2}
\end{bmatrix}\text{.}
\label{eqn:perspective3}
\end{equation}
The shift correction $\Delta_{x,\text{PCF}}$ along the $x_{\text{PCF}}$ axis is calculated on the PCF as follows	
\begin{equation}
\Delta_{x,\text{PCF}}=\frac{x_{i,\text{seen}}\cdot h_{i,\text{LTP}}}{h_{\text{UAV}}}\text{,}
\label{eqn:perspective2}
\end{equation}
where $h_{i,\text{LTP}}$ is the height of the \textit{i-th} corner on LTP and $h_{\text{UAV}}$ is the hovering altitude of the UAV in LTP. The correction $\Delta_{y,\text{PCF}}$ along the $y_{\text{PCF}}$ axis is calculated by analogy. The corrected coordinates $\boldsymbol{b}_{i,\text{c}}$ of $\boldsymbol{b}_i$ are then given by
\begin{equation}
\boldsymbol{b}_{i,\text{c}}=\boldsymbol{b}_i-\begin{bmatrix}
\Delta_{x,\text{PCF}}&
\Delta_{y,\text{PCF}}
\end{bmatrix}^\text{T}\text{.}
\label{eqn:perspective4}
\end{equation}
Since the detection error of $\pm\unit[1]{px}$ per axis is transferred to the bounding box, then $\boldsymbol{b}_i$ could have an error of $\unit[\sqrt{2}]{px}$. If this error occur towards the principal point, then $\Delta_{x,\text{PCF}}$ and $\Delta_{y,\text{PCF}}$ add an error towards the principal point as well. The error $\eta_{b,i}$ of $\boldsymbol{b}_i$ due to the shape detection and correction of the relief displacement is given by
\begin{equation}
\eta_{b,i}=\left(\sqrt{2}+\frac{h_{i,\text{LTP}}}{h_{\text{UAV}}}\right)\cdot\alpha\text{.}
\label{eqn:perspective5}
\end{equation}
As an example for a sedan with $\pm\unit[15]{cm}$ ground clearance and a UAV hovering at $\unit[100]{m}$, then $\eta_{b,i}\approx\unit[0.04]{m}$.

The corrected bounding box is reconstructed to find the centre of the vehicle. Let the width $w_{\text{car}}$ and length $l_{\text{car}}$ of the vehicle be known, the Equation (\ref{eqn:geomcentercalc2b}) hold, the spatial resolution have no errors, $\boldsymbol{b}_1$ be the corner closest to the image centre and its relief displacement be corrected. Then, $\boldsymbol{b}_1$ is used as base for scaling $\boldsymbol{b}_j$ and $\boldsymbol{b}_k$ as follows
\begin{equation}
\boldsymbol{b}_{j,\text{scaled}}=\left(\frac{l_{\text{car}}}{\boldsymbol{s}\left(2\right)\cdot\alpha}\cdot\left(\boldsymbol{b}_j-\boldsymbol{b}_1\right)\right)+\boldsymbol{b}_1\text{, and}
\label{eqn:boxbuild2}
\end{equation}
\begin{equation}
\boldsymbol{b}_{k,\text{scaled}}=\left(\frac{w_{\text{car}}}{\boldsymbol{s}\left(1\right)\cdot\alpha}\cdot\left(\boldsymbol{b}_k-\boldsymbol{b}_1\right)\right)+\boldsymbol{b}_1\text{,}
\label{eqn:boxbuild1}
\end{equation}
where $k$ is the element of $\boldsymbol{B}_{\text{PCF}}$ associated with $\boldsymbol{s}\left(1\right)$. The corrected centre of the vehicle is then calculated by
\begin{equation}
\boldsymbol{o}_{\text{in}}=\frac{\boldsymbol{b}_{j,\text{scaled}}+\boldsymbol{b}_{k,\text{scaled}}}{2}\text{.}
\label{eqn:boxbuild3}
\end{equation}
As stated before, the height cannot be easily determined for the two sides of the bounding box farthest from the vertical of the drone.
%This is particularly problematic for boxy vehicles, like off-roaders, where there the height can be 0.2 meters at one frame and 1.85 for the next. 
Considering passenger vehicles, this height can range from $h_{i,\text{min}}=\unit[0.11]{m}$ (ground clearance of sport vehicles~\cite{TUVRideHeight}\cite{RideHeightEU}) and up to $h_{i,\text{max}}=\unit[1.85]{m}$ (roof of multi-purpose vehicles~\cite{cartype}\cite{vehicleinertia}). If the relief displacement is not corrected, the maximum positioning error $\eta_{\text{scale}}$ due to wrong vehicle dimensions is given by
\begin{equation}
\eta_{\text{scale}}=\frac{\sqrt{x_{i,\text{seen}}^2+y_{i,\text{seen}}^2}\cdot\alpha\cdot\left(h_{i,\text{max}}-h_{i,\text{min}}\right)}{2\cdot h_{\text{UAV}}}\text{.}
\label{eqn:boxbuild4}
\end{equation}
For example, if the UAV hovers at $\unit[50]{m}$ and $\boldsymbol{b}_i$ is on one corner of the image, then $\eta_{\text{scale}}\approx\unit[0.63]{m}$.

%For works aiming for a general solution, a minimum ground clearance of 110 mm is suggested by the Verband der TÜV e.V.~\cite{TUVRideHeight}, and the European Union (EU) regulation No. 678/2011~\cite{RideHeightEU} classifies vehicles with ground clearance above 200 mm as off-road vehicles. An official EU vehicle classification can be found on~\cite{cartype}, and an extensive list of measured vehicle specifications can be found on~\cite{vehicleinertia}.

From the discussion above it follows that the best way to minimize positioning errors due to relief displacement, is to correct it only for the corner nearest to the centre of the image, and to re-scale the bounding box. If the vehicle measures are not known, generic values can be taken from the literature~\cite{KruberFuture}.
\subsection{Sensor Synchronization}\label{sec:SyncError}
Another relevant source of error is the synchronization of the sensors. A best practice approach is to use of the Pulse Per Second (PPS) from SatNav receivers and thus to employ the Universal Time Coordinated (UTC-time) as a common timeline. Atomic clocks aboard satellites~\cite{NASAClock} and negligible delays between receivers make this the best option. Unfortunately, most consumer-grade UAVs have no option for triggering the camera with a PPS pulse, nor do they provide a frame-accurate UTC timestamp. 

%An alternative for the PPS is the use of screens to display the UTC-time, and to record the screen with the camera; but this implies an unknown SatNav receiver-to-screen latency. Yet another alternative is the use of software triggers. However, since most consumer-grade UAVs run on Android-, Linux- or Windows-based operating systems, they are not real-time capable. Thus, there are unknown latencies between the software trigger, and the start of the video recording. These unknown latencies can be of several frames.

\begin{figure}%[htb]
	\vspace{2 mm}
	\centering
	\includegraphics[width=0.99\columnwidth]{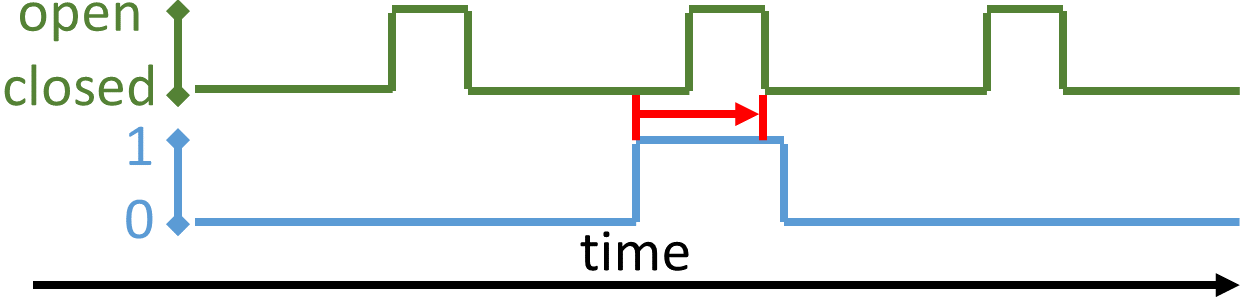}
	\caption{\small{Shown is: the PPS pulse from a SatNav receiver with blue, the shutter exposure time with green, and the synchronization error with red.}}
	\label{fig:LEDsync}
	\vspace{-0.5cm}
\end{figure}

In this work, the synchronization process is performed using as reference the light of the in-built LED of a Neo-6M~\cite{NeoGPS} SatNav receiver as follows. The rising edge of the PPS pulse indicates the start of every second. This rising edge is used as a trigger for lighting up a LED that stays on for a determined period of time so that the light can be seen on the image to be processed. Since this LED is part of the SatNav module, the latency between the PPS reception and LED lighting up is negligible. So, the start of each second can be known with frame-accuracy by recording this LED. A graphical representation of this is shown in \autoref{fig:LEDsync}. The limitation of this technique is that videos are not a continuous image, but a series of static pictures. So it is not known if the LED lights up when the shutter is closed, creating a synchronization error. The maximum synchronization error $\eta_{\tau}$ is given by  
\begin{equation}
\eta_{\tau}=\frac{1}{\tau_{\text{FR}}}\text{,}
\label{eqn:sync1}
\end{equation}
where $\tau_{\text{FR}}$ is the camera frame rate in frames per second (fps). In this work, $\unit[50]{fps}$ are used. So $\eta_{\tau}\leq\unit[.02]{s}$. The positioning error $\eta_{\text{pos}}$ caused by $\eta_{\tau}$ is given by
\begin{equation}
\eta_{\text{pos}}={\sqrt{v_{x,{\text{car}}}^2+v_{y,{\text{car}}}^2}}\cdot\eta_{\tau}\text{,}
\label{eqn:sync2}
\end{equation}
and the velocity error $\eta_{\text{vel}}$ caused by $\eta_{\tau}$ is given by
\begin{equation}
\eta_{\text{vel}}={\sqrt{a_{x,{\text{car}}}^2+a_{y,{\text{car}}}^2}}\cdot\eta_{\tau}\text{.}
\label{eqn:sync3}
\end{equation}
As an example with a vehicle moving with $\unitfrac[50]{km}{h}$, braking with $\unitfrac[5]{m}{s^2}$ and a UAV hovering at $\unit[100]{m}$ recording a video with $\unit[50]{fps}$, then $\eta_{\text{pos}}\leq\unit[0.25]{m}$ and $\eta_{\text{vel}}\leq\unit[.09]{m/s}$.

It can be deduced from what is discussed above that synchronization errors, even in the millisecond order, have a statistically significant influence. Also, if no PPS trigger is available, the best way to minimize errors due to synchronization is to film the on-board LED of SatNav receivers.
\section{Benchmark of the estimated vehicle state}\label{sec:results}
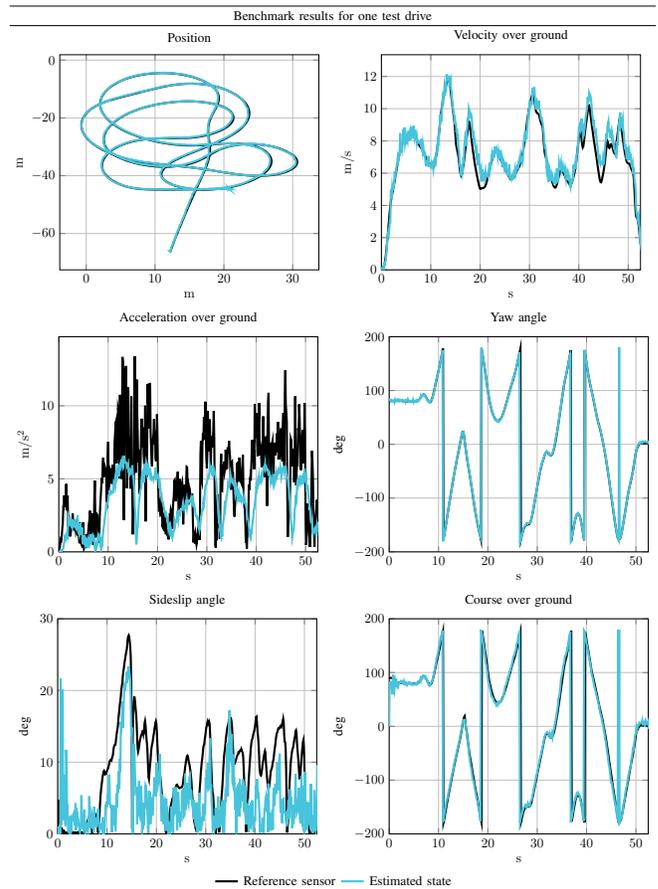
\begin{figure}
	\vspace{2 mm}
	\centering
	\renewcommand{\arraystretch}{1.2}
	\setlength{\tabcolsep}{2pt}
	\noindent
	\resizebox{0.49\textwidth}{!}{
		\begin{tabular}{cc}
			\hline
			\multicolumn{2}{c}{Benchmark results for one test drive}\\
			\hline
			\input{position1.tex} & \input{velocity1.tex} \\
			\input{acceleration1.tex} & \input{yaw1.tex} \\
			\input{sideslip1.tex} & \input{cog1.tex} \\
			\multicolumn{2}{c}{\normalsize{\ref{pl:Reference sensor} Reference sensor \ref{pl:Estimated state} Estimated state}}\\	
		\end{tabular}
	}
	\caption{\small{Shown are the benchmark results for one test drive with the UAV hovering at $\unit[75]{m}$.}}
	\label{fig:BEnchmark1Test}
	\vspace{-0.5cm}
\end{figure}	
The benchmark process is detailed in the following. The estimated state vector is compared to the outputs of an Automotive Dynamic Motion Analyzer (ADMA) G-Pro+~\cite{ADMAGenesys}. This Inertial Navigation System (INS) is equipped with servo-accelerometers, optical gyroscopes and receives Real-Time Kinematic (RTK) correction data. The estimated sideslip angle is compared to the output of a Correvit S-Motion~\cite{CorrevitKistler}. This sensor estimates velocities in the longitudinal and lateral axes by means of an optical grid~\cite{articleCorrevit}. The reference sideslip is computed analog to the Equation (\ref{eqn:KFEq1}). Finally, the UAV used is the DJI Phantom 4 Pro V2~\cite{DJIPhantom}.  

To generate the test data, a test vehicle is equipped with the reference sensors and is driven by a human in a random manner on a paved test track. No driving robot is used and no specific maneuver is driven in order to avoid tuning KF parameters to fit a specific trajectory, maneuver or driving style. The test drives include standstill, walking velocity, high acceleration, hard braking, tractive and non-tractive driving (drifting)~\cite{DrifitingDynamics}. The velocity is limited to \unit[50]{$\si{\kilo\meter\per\hour}$}. No markers are placed on the vehicle to approach real-testing conditions on open roads. To give variety to the test dataset, the videos are taken at three hovering altitudes: \unit[50]{m}, \unit[75]{m} and \unit[100]{m}. The videos are recorded with \unit[50]{fps} and 4K resolution (\unit[3840]{px}\,x\,\unit[2160]{px}), which are later compressed to FullHD (\unit[1920]{px}\,x\,\unit[1080]{px}). Four videos per height are recorded.

\autoref{fig:BEnchmark1Test} shows the benchmark results for one of the test drives with the biggest errors. For this trial run, the test vehicle is driven in such a manner to make it drift. This includes full-throttle acceleration, hard braking and sudden steering. %This test run is representative of the advantages and disadvantages of using camera-equipped UAVs as tools for automotive data acquisition. 

The estimated position, course over ground and yaw for this trial have a mean error of \unit[19.16]{cm}, $4.71^\circ$ and $1.02^\circ$ respectively. This precision is equivalent to that of consumer-grade INSs.

\autoref{fig:BEnchmark1Test} shows that the estimated velocity is affected by a dampening effect and by a time delay. Both are caused by the Kalman gains. A test-specific tuning of the gains could help to reduce the velocity error for this trial, but would increase the error for other tests with lower vehicle dynamics. 

The deviation of the estimated sideslip that is shown on \autoref{fig:BEnchmark1Test} is caused mainly by the velocity error. This is because the sideslip angle is estimated using the course over ground, which is calculated from the velocity (\autoref{sec:ResultsCompare}). During this test, a sideslip angle of $27.69^\circ$ is reached, which clearly indicates that the vehicle is drifting.

The error of the acceleration is explained by two facts: First, the estimated acceleration is calculated by system noise propagation (\autoref{sec:ResultsCompare}), so it is low-pass filtered. Second, the reference acceleration that is measured by the INS includes vibrations from the drivetrain, the tyres and the suspension, as well as from the pitch and roll of the vehicle.

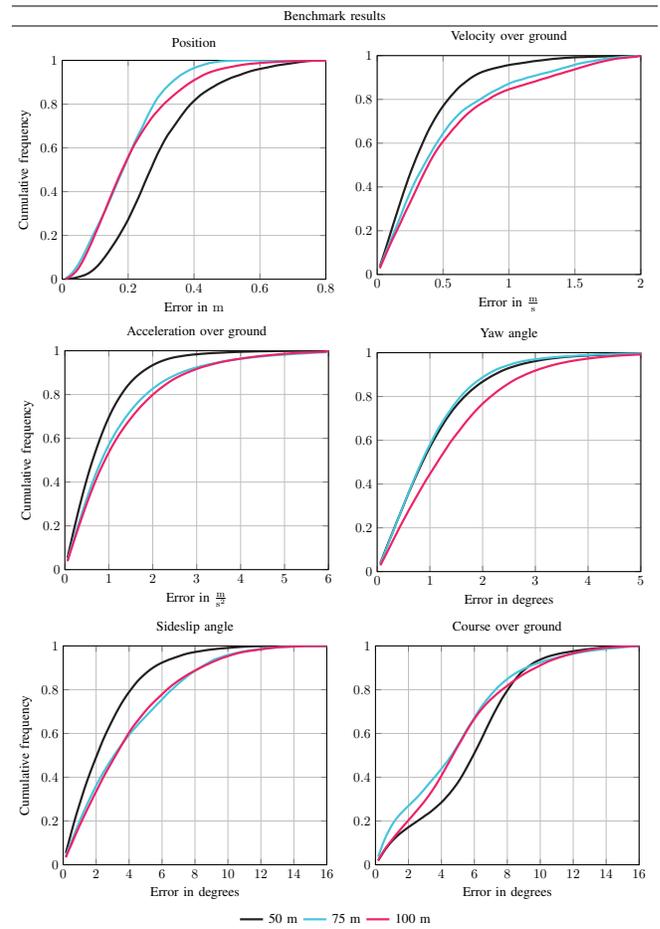
\begin{figure}
	\vspace{2 mm}
	\centering
	\renewcommand{\arraystretch}{1.2}
	\setlength{\tabcolsep}{2pt}
	\noindent
	\resizebox{0.49\textwidth}{!}{
		\begin{tabular}{cc}
			\hline
			\multicolumn{2}{c}{Benchmark results}\\
			\hline
			\input{position.tex} & \input{velocity.tex} \\
			\input{acceleration.tex} & \input{yaw.tex} \\
			\input{sideslip.tex} & \input{cog.tex} \\
			\multicolumn{2}{c}{\normalsize{\ref{pl:50mlabel} 50 m \ref{pl:75mlabel} 75 m \ref{pl:100mlabel} 100 m}}\\	
		\end{tabular}
	}
	\caption{\small{Shown are the cumulative frequency diagrams of the benchmark results for three hovering altitudes.}}
	\label{fig:BenchmarkAllTests}
	\vspace{-0.5cm}
\end{figure}

\autoref{fig:BenchmarkAllTests} shows the cumulative frequency diagrams for all performed tests. The benchmark results show that, once steps are taken to minimize errors, the precision of the estimated vehicle state is comparable to the precision of consumer-grade sensors, such as silicon-based INSs or SatNav receivers with no correction data. This with the advantage of being able to record information for various traffic participants with a single UAV. Also, the precision of the estimated sideslip angle allows to differentiate between tractive and non-tractive driving. This is a relevant state variable to determine the vehicle stability.
\section{Conclusions}\label{sec:Conclusions}
In this work, an evaluation of the accuracy of camera-equipped UAVs as tools for traffic data acquisition is presented. This includes a detailed description of the vehicle state estimation from aerial imagery, an analysis and quantification of the most relevant sources of error, and a benchmark of the estimated vehicle state with state-of-the-art reference sensors. The findings can then serve as a best-practice guide to minimize errors in future works. The source code is publicly available to aid researchers adapt KF gains for specific tasks.

The results of the benchmark show that, once steps are taken to minimize errors, camera-equipped UAVs are very attractive tools for automotive data acquisition. This because relatively low effort is required, while still obtaining a vehicle state with a precision comparable to that of consumer-grade INSs. This with the added benefit of recording various vehicles with a single device.

Being the correction of the relief displacement the source of error with the most unknowns, it has the biggest improving potential. If the height of the vehicle pixels is known in a deterministic manner, the positioning error can be greatly reduced.
\section{Acknowledgement}
The authors acknowledge the financial support by the Federal Ministry of Education and Research of Germany (BMBF) in the framework of FH-Impuls (project number 03FH7I02IA).
\bibliographystyle{IEEEtran}
\bibliography{literature}
\end{document}

%% file: architecture.tikz.tex
\tikzstyle{box}  = [shade,shading=axis,rectangle,draw,node distance=4mm and 4mm]%
\tikzstyle{boxb} = [box,rounded corners,top color=Cerulean,bottom color=MidnightBlue]%
\tikzstyle{boxg} = [box,rounded corners,top color=cyan,bottom color=cyan]%
\newcommand{\architecture}{%
    %\tikzpicturedependsonfile{fig/architecture.tikz.tex}%
    \node[boxg] (box0) {input};%
    \node[boxg] (box1) [right=of box0, align=center]{street data\\processing};%
    \node[boxg] (box2) [right=of box1, align=center]{trajectory\\generation};%
    \node[boxg] (box3) [right=of box2, align=center]{collision\\recognition};%
    \node[boxg] (box4) [right=of box3, align=center]{risk\\assessment};%
    \node[boxg] (box5) [right=of box4, align=center]{output};%
    \path[very thick,->] (box0) edge (box1);%
    \path[very thick,->] (box1) edge (box2);%
    \path[very thick,->] (box2) edge (box3);%
    \path[very thick,->] (box3) edge (box4);%
    \path[very thick,->] (box4) edge (box5);%
    \node[fit=(box1) (box4)] (box-cover) {};%
    \node[node distance=0mm and 0mm] (box-label) [above=of box-cover] {algorithm framework};%
    \begin{scope}[on background layer]
        \node[boxb, fit=(box-cover) (box-label)] (box-background) {};%
    \end{scope}
}%

%% file: position1.tex
\begin{tikzpicture}
%\footnotesize
\begin{axis}[
grid=major,
width=0.98\columnwidth,
%height=0.5\columnwidth,
xlabel=$\si{\meter}$,
ylabel=$\si{\meter}$,
%ymin=0, 
%ymax=1,
%xmin=0,
%xmax=0.8,
%xticklabels={2,4,6,8,10,12},
%xtick={2,4,6,8,10,12},
%xtick align=inside,
%              extra x ticks={2,4,6,8,10,12},
%      extra x tick labels={7,14,21,28,35,42},
%      every extra x tick/.style={major tick length=0pt,
%        xtick align=inside,ticklabel pos=top, color = paperBlue},
xlabel near ticks,
ylabel near ticks,
every axis plot/.append style={ultra thick},
legend pos=south east,
title= Position,
%every axis title/.style={below right,at={(0,1)},draw=black,fill=white}
]

\addplot[color=black] table [x=error,y=percent]{ADMAPositionForIV.dat};\label{pl:Reference sensor}
%\addlegendentry{50m}
\addplot[color=SkyBlue] table [x=error,y=percent]{DronePositionForIV.dat};\label{pl:Estimated state}
%\addlegendentry{100m}
\end{axis}
\end{tikzpicture}

%% file: velocity1.tex
\begin{tikzpicture}
%\footnotesize
\begin{axis}[
grid=major,
width=0.98\columnwidth,
%height=0.5\columnwidth,
xlabel=$\si{\second}$,
ylabel=${\si{\meter}}/{\si{\second}}$,
ymin=0, 
xmin=0,
xmax=52.5,
%xticklabels={2,4,6,8,10,12},
%xtick={2,4,6,8,10,12},
%xtick align=inside,
%              extra x ticks={2,4,6,8,10,12},
%      extra x tick labels={7,14,21,28,35,42},
%      every extra x tick/.style={major tick length=0pt,
%        xtick align=inside,ticklabel pos=top, color = paperBlue},
xlabel near ticks,
ylabel near ticks,
every axis plot/.append style={ultra thick},
legend pos=south east,
title= Velocity over ground,
%every axis title/.style={below right,at={(0,1)},draw=black,fill=white}
]

\addplot[color=black] table [x=error,y=percent]{ADMAVelocityForIV.dat};%\label{pl:50mlabel}
%\addlegendentry{50m}
\addplot[color=SkyBlue] table [x=error,y=percent]{DroneVelocityForIV.dat};%\label{pl:75mlabel}
%\addlegendentry{75m}
\end{axis}
\end{tikzpicture}

%% file: acceleration1.tex
\begin{tikzpicture}
%\footnotesize
\begin{axis}[
grid=major,
width=0.98\columnwidth,
%height=0.5\columnwidth,
xlabel=$\si{\second}$,
ylabel=${\si{\meter}}/{\si{\second}}^2$,
ymin=0, 
xmin=0,
xmax=52.5,
%xticklabels={2,4,6,8,10,12},
%xtick={2,4,6,8,10,12},
%xtick align=inside,
%              extra x ticks={2,4,6,8,10,12},
%      extra x tick labels={7,14,21,28,35,42},
%      every extra x tick/.style={major tick length=0pt,
%        xtick align=inside,ticklabel pos=top, color = paperBlue},
xlabel near ticks,
ylabel near ticks,
every axis plot/.append style={ultra thick},
legend pos=south east,
title= Acceleration over ground,
%every axis title/.style={below right,at={(0,1)},draw=black,fill=white}
]

\addplot[color=black] table [x=error,y=percent]{ADMAAccelerationForIV.dat};%\label{pl:50mlabel}
%\addlegendentry{50m}
\addplot[color=SkyBlue] table [x=error,y=percent]{DroneAccelerationForIV.dat};%\label{pl:75mlabel}
%\addlegendentry{75m}
\end{axis}
\end{tikzpicture}

%% file: yaw1.tex
\begin{tikzpicture}
%\footnotesize
\begin{axis}[
grid=major,
width=0.98\columnwidth,
%height=0.5\columnwidth,
xlabel=$\si{\second}$,
ylabel=deg,
ymin=-200, 
ymax=200,
xmin=0,
xmax=52.5,
%xticklabels={2,4,6,8,10,12},
%xtick={2,4,6,8,10,12},
%xtick align=inside,
%              extra x ticks={2,4,6,8,10,12},
%      extra x tick labels={7,14,21,28,35,42},
%      every extra x tick/.style={major tick length=0pt,
%        xtick align=inside,ticklabel pos=top, color = paperBlue},
xlabel near ticks,
ylabel near ticks,
every axis plot/.append style={ultra thick},
legend pos=south east,
title= Yaw angle,
%every axis title/.style={below right,at={(0,1)},draw=black,fill=white}
]

\addplot[color=black] table [x=error,y=percent]{ADMAYawForIV.dat};%\label{pl:50mlabel}
%\addlegendentry{50m}
\addplot[color=SkyBlue] table [x=error,y=percent]{DroneYawForIV.dat};%\label{pl:75mlabel}
%\addlegendentry{75m}
\end{axis}
\end{tikzpicture}

%% file: sideslip1.tex
\begin{tikzpicture}
%\footnotesize
\begin{axis}[
grid=major,
width=0.98\columnwidth,
%height=0.5\columnwidth,
xlabel=$\si{\second}$,
ylabel=deg,
ymin=0, 
ymax=30,
xmin=0,
xmax=52.5,
%xticklabels={2,4,6,8,10,12},
%xtick={2,4,6,8,10,12},
%xtick align=inside,
%              extra x ticks={2,4,6,8,10,12},
%      extra x tick labels={7,14,21,28,35,42},
%      every extra x tick/.style={major tick length=0pt,
%        xtick align=inside,ticklabel pos=top, color = paperBlue},
xlabel near ticks,
ylabel near ticks,
every axis plot/.append style={ultra thick},
legend pos=south east,
title= Sideslip angle,
%every axis title/.style={below right,at={(0,1)},draw=black,fill=white}
]

\addplot[color=black] table [x=error,y=percent]{ADMASideslipForIV.dat};%\label{pl:regshiftfjdf}
%\addlegendentry{50m}
\addplot[color=SkyBlue] table [x=error,y=percent]{DroneSideslipForIV.dat};%\label{pl:xxff}
%\addlegendentry{75m}
\end{axis}
\end{tikzpicture}

%% file: cog1.tex
\begin{tikzpicture}
%\footnotesize
\begin{axis}[
grid=major,
width=0.98\columnwidth,
%height=0.5\columnwidth,
xlabel=$\si{\second}$,
ylabel=deg,
ymin=-200, 
ymax=200,
xmin=0,
xmax=52.5,
%xticklabels={2,4,6,8,10,12},
%xtick={2,4,6,8,10,12},
%xtick align=inside,
%              extra x ticks={2,4,6,8,10,12},
%      extra x tick labels={7,14,21,28,35,42},
%      every extra x tick/.style={major tick length=0pt,
%        xtick align=inside,ticklabel pos=top, color = paperBlue},
xlabel near ticks,
ylabel near ticks,
every axis plot/.append style={ultra thick},
legend pos=south east,
title= Course over ground,
%every axis title/.style={below right,at={(0,1)},draw=black,fill=white}
]

\addplot[color=black] table [x=error,y=percent]{ADMACOGForIV.dat};%\label{pl:50mlabel}
%\addlegendentry{50m}
\addplot[color=SkyBlue] table [x=error,y=percent]{DroneCOGForIV.dat};%\label{pl:75mlabel}
%\addlegendentry{75m}
\end{axis}
\end{tikzpicture}

%% file: position.tex
\begin{tikzpicture}
%\footnotesize
\begin{axis}[
grid=major,
width=0.98\columnwidth,
%height=0.5\columnwidth,
xlabel=Error in $\si{\meter}$,
ylabel=Cumulative frequency,
ymin=0, 
ymax=1,
xmin=0,
xmax=0.8,
%xticklabels={2,4,6,8,10,12},
%xtick={2,4,6,8,10,12},
%xtick align=inside,
%              extra x ticks={2,4,6,8,10,12},
%      extra x tick labels={7,14,21,28,35,42},
%      every extra x tick/.style={major tick length=0pt,
%        xtick align=inside,ticklabel pos=top, color = paperBlue},
xlabel near ticks,
ylabel near ticks,
every axis plot/.append style={ultra thick},
legend pos=south east,
title= Position,
%every axis title/.style={below right,at={(0,1)},draw=black,fill=white}
]

\addplot[color=Black] table [x=error,y=percent]{AAxesPosition50.dat};\label{pl:50mlabel}
%\addlegendentry{50m}
\addplot[color=SkyBlue] table [x=error,y=percent]{AAxesPosition75.dat};\label{pl:75mlabel}
%\addlegendentry{75m}
\addplot[color=WildStrawberry] table [x=error,y=percent]{AAxesPosition100.dat};\label{pl:100mlabel}
%\addlegendentry{100m}
\end{axis}
\end{tikzpicture}

%% file: velocity.tex
\begin{tikzpicture}
%\footnotesize
\begin{axis}[
grid=major,
width=0.98\columnwidth,
%height=0.5\columnwidth,
xlabel=Error in $\frac{\si{\meter}}{\si{\second}}$,
%ylabel=Cumulative frequency,
ymin=0, 
ymax=1,
xmin=0,
xmax=2,
%xticklabels={2,4,6,8,10,12},
%xtick={2,4,6,8,10,12},
%xtick align=inside,
%              extra x ticks={2,4,6,8,10,12},
%      extra x tick labels={7,14,21,28,35,42},
%      every extra x tick/.style={major tick length=0pt,
%        xtick align=inside,ticklabel pos=top, color = paperBlue},
xlabel near ticks,
ylabel near ticks,
every axis plot/.append style={ultra thick},
legend pos=south east,
title= Velocity over ground,
%every axis title/.style={below right,at={(0,1)},draw=black,fill=white}
]

\addplot[color=Black] table [x=error,y=percent]{AAxesVelocity50.dat};%\label{pl:50mlabel}
%\addlegendentry{50m}
\addplot[color=SkyBlue] table [x=error,y=percent]{AAxesVelocity75.dat};%\label{pl:75mlabel}
%\addlegendentry{75m}
\addplot[color=WildStrawberry] table [x=error,y=percent]{AAxesVelocity100.dat};%\label{pl:100mlabel}
%\addlegendentry{100m}
\end{axis}
\end{tikzpicture}

%% file: acceleration.tex
\begin{tikzpicture}
%\footnotesize
\begin{axis}[
grid=major,
width=0.98\columnwidth,
%height=0.5\columnwidth,
xlabel=Error in $\frac{\si{\meter}}{\si{\second}^2}$,
ylabel=Cumulative frequency,
ymin=0, 
ymax=1,
xmin=0,
xmax=6,
%xticklabels={2,4,6,8,10,12},
%xtick={2,4,6,8,10,12},
%xtick align=inside,
%              extra x ticks={2,4,6,8,10,12},
%      extra x tick labels={7,14,21,28,35,42},
%      every extra x tick/.style={major tick length=0pt,
%        xtick align=inside,ticklabel pos=top, color = paperBlue},
xlabel near ticks,
ylabel near ticks,
every axis plot/.append style={ultra thick},
legend pos=south east,
title= Acceleration over ground,
%every axis title/.style={below right,at={(0,1)},draw=black,fill=white}
]

\addplot[color=Black] table [x=error,y=percent]{AAxesAcceleration50.dat};%\label{pl:50mlabel}
%\addlegendentry{50m}
\addplot[color=SkyBlue] table [x=error,y=percent]{AAxesAcceleration75.dat};%\label{pl:75mlabel}
%\addlegendentry{75m}
\addplot[color=WildStrawberry] table [x=error,y=percent]{AAxesAcceleration100.dat};%\label{pl:100mlabel}
%\addlegendentry{100m}
\end{axis}
\end{tikzpicture}

%% file: yaw.tex
\begin{tikzpicture}
%\footnotesize
\begin{axis}[
grid=major,
width=0.98\columnwidth,
%height=0.5\columnwidth,
xlabel=Error in degrees,
%ylabel=Cumulative frequency,
ymin=0, 
ymax=1,
xmin=0,
xmax=5,
%xticklabels={2,4,6,8,10,12},
%xtick={2,4,6,8,10,12},
%xtick align=inside,
%              extra x ticks={2,4,6,8,10,12},
%      extra x tick labels={7,14,21,28,35,42},
%      every extra x tick/.style={major tick length=0pt,
%        xtick align=inside,ticklabel pos=top, color = paperBlue},
xlabel near ticks,
ylabel near ticks,
every axis plot/.append style={ultra thick},
legend pos=south east,
title= Yaw angle,
%every axis title/.style={below right,at={(0,1)},draw=black,fill=white}
]

\addplot[color=Black] table [x=error,y=percent]{AAxesYaw50.dat};%\label{pl:50mlabel}
%\addlegendentry{50m}
\addplot[color=SkyBlue] table [x=error,y=percent]{AAxesYaw75.dat};%\label{pl:75mlabel}
%\addlegendentry{75m}
\addplot[color=WildStrawberry] table [x=error,y=percent]{AAxesYaw100.dat};%\label{pl:100mlabel}
%\addlegendentry{100m}
\end{axis}
\end{tikzpicture}

%% file: sideslip.tex
\begin{tikzpicture}
%\footnotesize
\begin{axis}[
grid=major,
width=0.98\columnwidth,
%height=0.5\columnwidth,
xlabel=Error in degrees,
ylabel=Cumulative frequency,
ymin=0, 
ymax=1,
xmin=0,
xmax=16,
%xticklabels={2,4,6,8,10,12},
%xtick={2,4,6,8,10,12},
%xtick align=inside,
%              extra x ticks={2,4,6,8,10,12},
%      extra x tick labels={7,14,21,28,35,42},
%      every extra x tick/.style={major tick length=0pt,
%        xtick align=inside,ticklabel pos=top, color = paperBlue},
xlabel near ticks,
ylabel near ticks,
every axis plot/.append style={ultra thick},
legend pos=south east,
title= Sideslip angle,
%every axis title/.style={below right,at={(0,1)},draw=black,fill=white}
]

\addplot[color=Black] table [x=error,y=percent]{AAxesSideSlip50.dat};%\label{pl:regshiftfjdf}
%\addlegendentry{50m}
\addplot[color=SkyBlue] table [x=error,y=percent]{AAxesSideSlip75.dat};%\label{pl:xxff}
%\addlegendentry{75m}
\addplot[color=WildStrawberry] table [x=error,y=percent]{AAxesSideSlip100.dat};%\label{pl:xxxdfe}
%\addlegendentry{100m}
\end{axis}
\end{tikzpicture}

%% file: cog.tex
\begin{tikzpicture}
%\footnotesize
\begin{axis}[
grid=major,
width=0.98\columnwidth,
%height=0.5\columnwidth,
xlabel=Error in degrees,
%ylabel=Cumulative frequency,
ymin=0, 
ymax=1,
xmin=0,
xmax=16,
%xticklabels={2,4,6,8,10,12},
%xtick={2,4,6,8,10,12},
%xtick align=inside,
%              extra x ticks={2,4,6,8,10,12},
%      extra x tick labels={7,14,21,28,35,42},
%      every extra x tick/.style={major tick length=0pt,
%        xtick align=inside,ticklabel pos=top, color = paperBlue},
xlabel near ticks,
ylabel near ticks,
every axis plot/.append style={ultra thick},
legend pos=south east,
title= Course over ground,
%every axis title/.style={below right,at={(0,1)},draw=black,fill=white}
]

\addplot[color=Black] table [x=error,y=percent]{AAxesCOG50.dat};%\label{pl:50mlabel}
%\addlegendentry{50m}
\addplot[color=SkyBlue] table [x=error,y=percent]{AAxesCOG75.dat};%\label{pl:75mlabel}
%\addlegendentry{75m}
\addplot[color=WildStrawberry] table [x=error,y=percent]{AAxesCOG100.dat};%\label{pl:100mlabel}
%\addlegendentry{100m}
\end{axis}
\end{tikzpicture}